\documentclass{iau}
\usepackage{graphicx}

\title[LEECH] 
{LEECH: A 100 Night Exoplanet Imaging Survey at the LBT}

\author[Andrew Skemer]   
{Andrew Skemer$^1$,
Daniel Apai$^1$,
Vanessa Bailey$^1$,
Beth Biller$^2$,
Mickael Bonnefoy$^2$,
Wolfgang Brandner$^2$,
Esther Buenzli$^2$,
Laird Close$^1$,
Justin Crepp$^3$,
Denis Defrere$^1$,
Silvano Desidera$^4$,
Josh Eisner$^1$,
Simone Esposito$^5$,
Jonathan Fortney$^6$,
Thomas Henning$^2$,
Phil Hinz$^1$,
Karl-Heinz Hofmann$^7$,
Jarron Leisenring$^{1,8}$,
Jared Males$^1$,
Rafael Millan-Gabet$^9$,
Katie Morzinski$^1$,
Apurva Oza$^{10}$,
Ilaria Pascucci$^{11}$,
Jenny Patience$^{12}$,
George Rieke$^1$,
Dieter Schertl$^7$,
Joshua Schlieder$^2$,
Mike Skrutskie$^{10}$,
Kate Su$^1$,
Gerd Weigelt$^7$,
Charles E. Woodward$^{13}$,
\and Neil Zimmerman$^2$}

\affiliation{$^1$Steward Observatory, Department of Astronomy, University of Arizona\\
$^2$Max Planck Institute for Astronomy\\
$^3$Department of Physics, Notre Dame University\\
$^4$Osservatorio Astronomico di Padova, Istituto Nazionale di Astrofisica\\
$^5$Osservatorio Astrofisico di Arcetri, Istituto Nazionale di Astrofisica\\
$^6$Department of Astronomy and Astrophysics, University of California, Santa Cruz\\
$^7$Max Planck Institute for Radio Astronomy\\
$^8$Institute for Astronomy, Eldgenossische Technische Hochschule Zurich\\
$^9$NASA Exoplanet Science Institute, California Institute of Technology\\
$^{10}$Department of Astronomy, University of Virginia\\
$^{11}$Department of Planetary Sciences, Lunar and Planetary Laboratory, University of Arizona\\
$^{12}$School of Earth and Space Exploration, Arizona State University\\
$^{13}$Minnesota Institute for Astrophysics, University of Minnesota}

\pubyear{2013}
\volume{299}  
\pagerange{xxx--xxx}
\jname{Exploring the Formation and Evolution of Planetary Systems}
\editors{Brenda Matthews \& James Graham, eds.}
\begin{document}

\maketitle

\begin{abstract}
In February 2013, the LEECH (LBTI Exozodi Exoplanet Common Hunt) survey began its 100-night campaign from the Large Binocular Telescope atop Mount Graham in Arizona. LEECH neatly complements other high-contrast planet imaging efforts by observing stars in L' band (3.8 microns) as opposed to the shorter wavelength near-infrared bands (1--2.3 microns). This part of the spectrum offers deeper mass sensitivity for intermediate age (several hundred Myr-old) systems, since their Jovian-mass planets radiate predominantly in the mid-infrared.  In this proceedings, we present the science goals for LEECH and a preliminary contrast curve from some early data.
\end{abstract}


\begin{description}
\item[Discovering Adolescent Exoplanets:]
\hspace{3pt}
Most exoplanet imaging surveys operate in the near-infrared (H-band; 1.65 $\mu$m), which provides the best contrasts at small separations.  However, self-luminous exoplanets emit most of their light at longer wavelengths, and this becomes more important as exoplanets age and lose the residual energy from their formation.  By operating in the mid-infrared (L’; 3.8 $\mu$m) LBTI is sensitive to planets around older stars (see Figure 1).  This will allow LEECH to discover “adolescent” exoplanets around nearby stars, complementing the young samples discovered by other surveys (see Figure 2), such as GPI, SPHERE, SEEDS, and Project 1640 (Macintosh et al. 2008, Beuzit et al. 2008, Tamura 2009, Hinkley et al. 2011)
\item[Connecting Planets with Disks:] 
\hspace{3pt}
Using the two LBTI science cameras in parallel, LEECH and HOSTS (see paper by Defrere et al. in these proceedings) will search for giant planets and inner debris disks simultaneously.  The LEECH and HOSTS samples include very nearby stars that have been targeted by other relevant surveys (Doppler-RV planet searches and outer debris disk studies).  Combining this data, will provide the first comprehensive view of exoplanetary systems.
\item[Characterizing Exoplanet Atmospheres:]
\hspace{3pt}
Early observations suggest that directly imaged planets have different colors and spectra than their field brown-dwarf analogs.  By studying planets with LBTI's broad wavelength coverage (1-13$\mu$m) and over a wide age range, LEECH will characterize the bulk atmospheric properties and evolution of extrasolar planets. 
\end{description}

\vspace{-1pt}

\begin{figure}[h]
\begin{center}
 \includegraphics[width=4.3in]{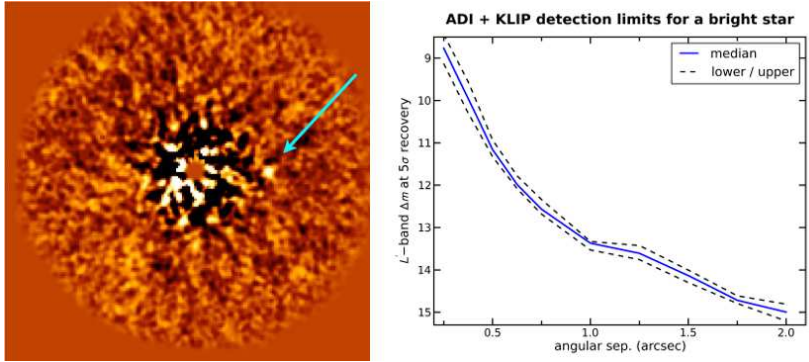} 
 \caption{LEFT: A first light image from LEECH with the star removed and an artificial planet inserted.  The artificial planet is 13 magnitudes fainter than the star at a separation of 0.75'', equivalent to a 3 Mjup planet 7.5 AU from a 0.5 Gyr solar type star at 10 pc.  LEECH's ability to image older exoplanets than other surveys will extend our knowledge of exoplanet evolution. RIGHT: ADI+PCA (KLIP) detection limits for a bright star with 2 hours integration time on 1 telescope (preliminary).
}
   \label{fig1}
\end{center}
\end{figure}

\vspace{-18pt}

\begin{figure}[h]
\begin{center}
 \includegraphics[width=4.3in]{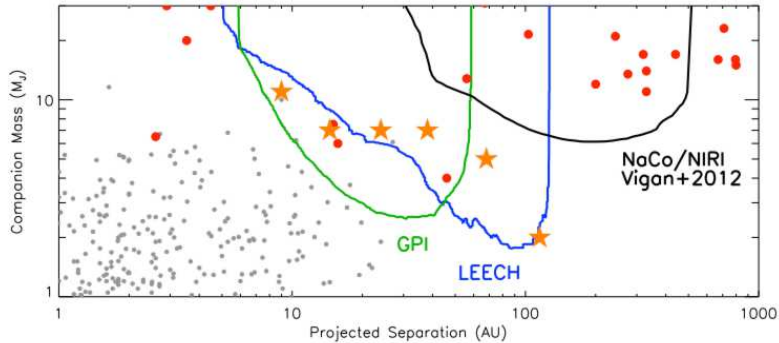} 
\vspace{-10pt}
 \caption{Planet mass versus semi-major axis for the known exoplanets and planetary mass companions listed in www.exoplanet.eu shown as small circles for radial velocity detections, large circles for imaging, and stars for imaged planets around HR 8799, $\beta$  Pic, and Fomalhaut. The median planet mass sensitivities are shown for the GPI A- and F-star sample, the LEECH sample, and an A-star search (Vigan et al. 2012) with NaCo/NIRI. Both GPI and LEECH will explore the critical missing link between radial velocity searches and current AO imaging surveys.
}
   \label{fig2}
\end{center}
\end{figure}

\end{document}